\documentclass[journal]{IEEEtran}

\usepackage{graphicx}
\usepackage{epsfig}
\usepackage{dcolumn}
\usepackage{bm}
\usepackage{amsmath}
\usepackage{amssymb}

\begin{document}
%
\title{Analog-to-Digital and Digital-to-Analog Conversion with Memristive Devices}
%
%
%

\author{Yuriy~V.~Pershin, Edward~Sazonov and Massimiliano~Di~Ventra
\thanks{Y. V. Pershin is with the Department of Physics
and Astronomy and USC Nanocenter, University of South Carolina,
Columbia, SC, 29208 \newline e-mail: pershin@physics.sc.edu}
\thanks{E. Sazonov is with the Department of Electrical and Computer Engineering, University of Alabama,
 Tuscaloosa, AL 35487 \newline e-mail: esazonov@eng.ua.edu}
\thanks{M. Di Ventra is with the Department
of Physics, University of California, San Diego, La Jolla,
CA 92093; e-mail: diventra@physics.ucsd.edu}
}

%
%


\maketitle

\begin{abstract}
We suggest a novel methodology to obtain a digital representation of analog signals and to perform its back-conversion using memristive devices. In the proposed converters, the same memristive systems are used for two purposes: as elements performing conversion and elements storing the code. This approach to conversion is particularly relevant for interfacing analog signals with memristive digital logic/computing circuits.
\end{abstract}

\begin{IEEEkeywords}
Analog-digital conversion, Digital-analog conversion, Memristors, Memcapacitors, Meminductors
\end{IEEEkeywords}

%
\IEEEpeerreviewmaketitle

\IEEEPARstart{A}{nalog}-to-digital converters (ADC) are widely used in modern electronics to represent values of analog signals in digital form for subsequent storage and/or logic processing by traditional micro-controllers. Digital-to-analog converters (DAC) perform the inverse operation, by transforming the digital information into the analog form. Recently, a different {\it memristive} logic architecture has been demonstrated \cite{borghetti10a,pershin10e} in which the same memristive devices \cite{chua76a} play the role of gate and latch. In this paper, we suggest an approach to analog-to-digital (and digital-to-analog) conversion with memristive devices in which the digital value of the signal is written into (read out of) the states of memristive systems. This approach is thus directly compatible with the {\it memristive} logic architectures recently proposed \cite{borghetti10a,pershin10e} and can also be used in chips interfacing the usual electronic components. Even though we do not present it here, we
anticipate that the methodology we suggest could be also implemented with memcapacitive and meminductive elements (capacitors and inductors with memory, respectively),~\cite{diventra09a} which, in principle, can be made almost dissipationless.

\begin{figure*}[tp]
 \begin{center}
\includegraphics[width=13cm]{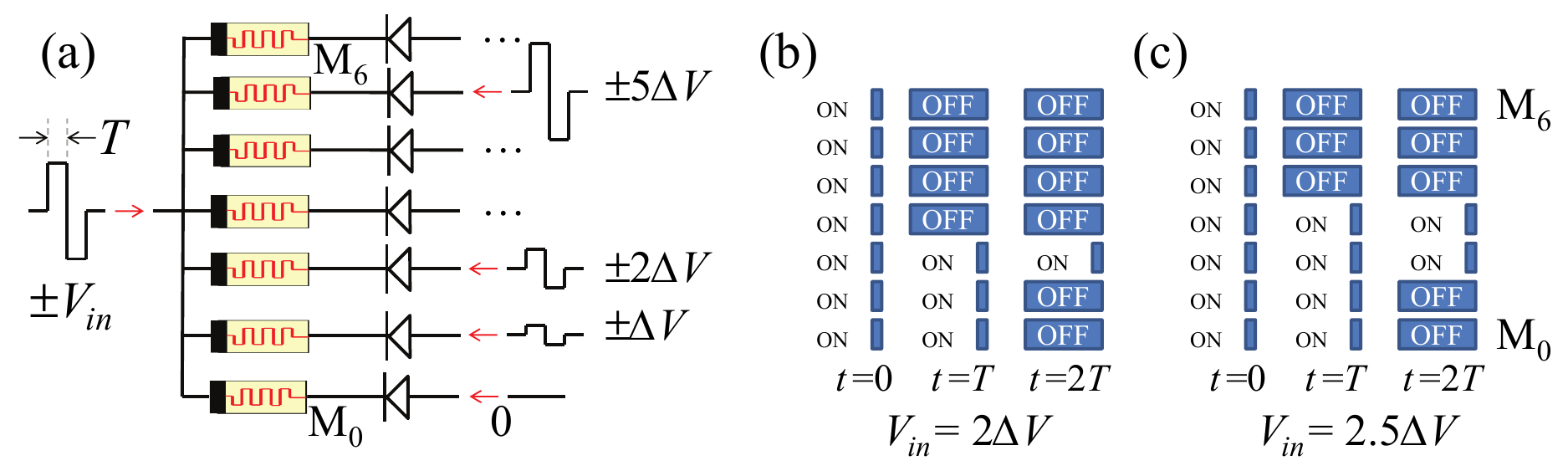}
\caption{(a) Schematic of a unipolar analog-to-digital converter based on memristive systems with threshold voltage $V_{th}=3/4 \Delta V$. Its operation involves a simultaneous application of double-step square pulses of indicated magnitude. The diodes are used in order to avoid 'OFF' to 'ON' switchings when the negative part of double-step square pulses is applied. The states of memristive systems at different selected moments of time are shown in (b) and (c) for two representative values of the input voltage. The order of states in (b) and (c) corresponds to the order of memristive systems in (a). \label{fig1}}
 \end{center}
\end{figure*}

Our approach requires the use of bipolar memristive systems with a threshold-type switching. Currently, these are the most studied types of memristive systems important for non-volatile information storage applications~\cite{pershin11a}. Physically, it is reasonable to think that most processes that lead to memory require a threshold energy to be activated. A voltage-controlled memristive system (belonging to the general class of memory circuit elements \cite{diventra09a}) is defined by the following equations \cite{chua76a,diventra09a}:
\begin{eqnarray}
I_M&=&\left[ R_M\left( x,V_M,t \right)\right]^{-1}V_M, \label{RM1}  \\
\frac{\textnormal{d}x}{\textnormal{d}t}&=&f\left( x,V_M,t \right), \label{RM2}
\end{eqnarray}
where $I_M$ and $V_M$ are the current through and the voltage across the memristive system, respectively \cite{chua76a}, $R_M$ is the memristance (memory resistance), $x$ is a vector of internal state variables, and $f$ is a device-specific function. In this work we will consider memristive memory cells (that can be based on electrochemical metallization or valency change mechanism)~\cite{pershin11a} exhibiting a threshold-type bipolar switching defined by a threshold voltage $V_{th}$ and limiting values of memristance, $R_{ON}$ and $R_{OFF}$ ($R_{ON}\ll R_{OFF}$). While the particular forms of $f\left( x,V_M,t \right)$ and $R_M\left( x,V_M,t \right)$ are not important for our consideration, we emphasize that we require that $R_M$ changes only when $V_M>V_{th}$.

The main idea behind ADC implementation is shown in Fig. \ref{fig1}(a). The analog-to-digital conversion is performed when synchronized double-step pulses are simultaneously applied to a set of $N$ memristive systems initially prepared in the 'ON' (low-resistance) state (also denoted by '1'). The duration of the single pulse, $T$, is selected in such a way that if the voltage across any memristive system in the set exceeds the threshold voltage $V_{th}$ then such memristive system switches into the 'OFF' (high-resistance) state (also denoted by '0'). The amplitude of pulses applied from the right of Fig. \ref{fig1}(a) is $i \Delta V$, where $i$ is the index of the memristive system in the set, $i=0,..., N-1$ and $\Delta V=(4/3)V_{th}$. This value of $\Delta V$ is specifically selected in order to obtain the same input voltage interval per each ADC code. Note, that we avoid selecting $\Delta V=V_{th}/2$ to avoid missing or incorrect code value when the input signal is close to a voltage threshold separating different ADC codes. The double pulse applied from the left is defined by the input voltage amplitude $V_{in} \ge 0$.

The first positive pulse switches all memristive systems with indexes $i>(V_{in}+V_{th})/\Delta V$ into the 'OFF' state. The second negative pulse does the same with all memristive systems with $i<(V_{in}-V_{th})/\Delta V$. It is clear that the applied set of pulses leaves only one or two memristive systems in the 'ON' state. This depends on whether $V_{in}$ is close to $j\Delta V$ or to $(j+0.5)\Delta V$ where $j$ is an integer number. Figs. \ref{fig1}(b,c) show main steps of switching dynamics in both cases. Table \ref{table_adc} provides the correspondence between the input voltage intervals and ADC output code. Note also that it is not necessary to map analog values into all binary numbers: using elementary logic operations with memristive devices \cite{pershin10e}, any desired representation of the output code can be obtained.

\begin{table}[!t]
\renewcommand{\arraystretch}{1.3}
\caption{ADC output code for different ranges of input signal}
\label{table_adc}
\centering
\begin{tabular}{c|c}
\hline
Input signal range & Code\\
\hline
$0 \le V_{in} < 0.25 \Delta V $ & 0000001 \\
$0.25 \Delta V \le V_{in} < 0.75 \Delta V $ & 0000011 \\
$0.75 \Delta V \le V_{in} < 1.25 \Delta V $ & 0000010 \\
$1.25 \Delta V \le V_{in} < 1.75 \Delta V $ & 0000110 \\
$1.75 \Delta V \le V_{in} < 2.25 \Delta V $ & 0000100 \\
$2.25 \Delta V \le V_{in} < 2.75 \Delta V $ & 0001100 \\
$2.75 \Delta V \le V_{in} < 3.25 \Delta V $ & 0001000 \\
... & ... \\
\hline
\end{tabular}
\end{table}

\begin{figure}[b]
 \begin{center}
\includegraphics[width=4.5cm]{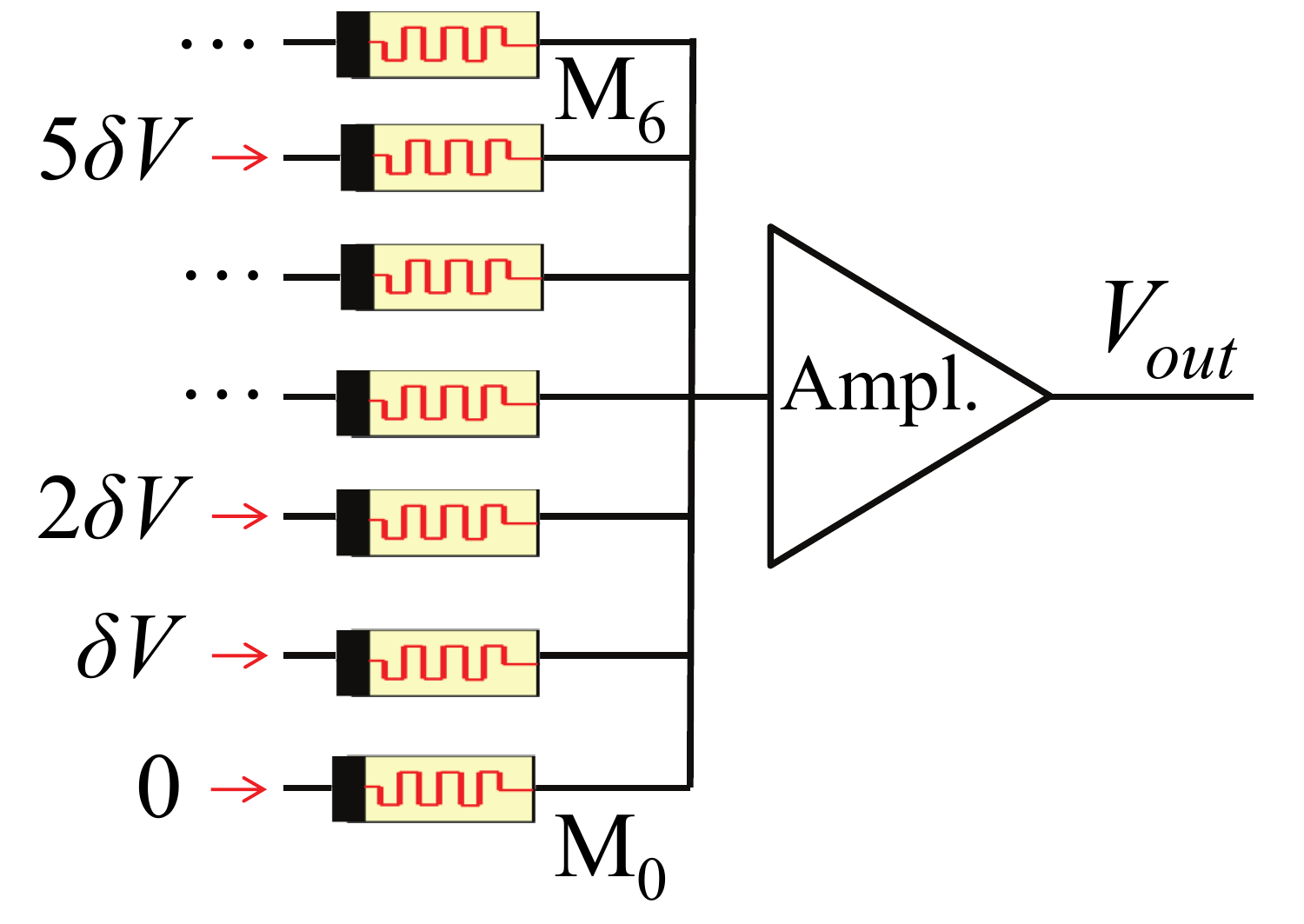}
\caption{Schematic of a digital-to-analog converter. Note that only one or two memristive systems can be in the 'ON' state specified by $R_{ON}\ll R_{OFF}$. The choice of $\delta V$ and amplifier's gain is described in the text. \label{fig2}}
 \end{center}
\end{figure}

The scheme shown in Fig. \ref{fig2} is used as DAC. For its correct operation, it is important to apply small amplitude signals to memristive systems in order to keep their states unchanged \cite{pershin10d}. Therefore, the maximum allowable $\delta V$ is determined from the relation $(N-1)\delta V<V_{th}$. In order to restore the initial signal magnitude, the gain of the amplifier schematically shown to the right of Fig. \ref{fig2} should be set to $\gamma=\Delta V/\delta V$.

In conclusion, we have presented electronic circuits performing analog-to-digital and digital-to-analog conversions. These circuits are based on memristive devices and thus can find widespread use, for example, as interfaces between analog circuits with logic/computing circuits based on memristors. We note that the conversion time of memristive ADCs is of the order of twice the switching time of the memristive system. For memristive memory cells, the switching time can be as short as 5ns \cite{pershin11a} resulting in 0.1Ghz conversion frequency. Experimentally, small threshold voltages $\sim 0.1$V have been observed in certain solid state memristive systems~\cite{Dietrich07a}. Therefore, for a $V=5$V input range, the  memristive ADC can distinguish $2V/\Delta V\approx 75$ voltage levels. This is more than enough for such applications as, e.g., multi-level or fuzzy logic. We also anticipate that memcapacitive or meminductive systems can be alternatively employed in ADCs and DACs, with the added advantage that they can be chosen, in principle, almost dissipationless.

\bibliographystyle{IEEEtran}
\bibliography{IEEEabrv,maze}

\end{document}